\documentclass[twocolumn]{revtex4}
\usepackage{amssymb}
\usepackage{latexsym}
\usepackage{epsfig}

\usepackage{amsmath}
\begin{document}
\title{ Diagnosing  the R$\acute{e}$nyi Holographic Dark Energy model in a flat Universe}
\author{ Vipin Chandra Dubey$^{1}$\footnote{ vipin.dubey@gla.ac.in} Ambuj Kumar Mishra$^{2}$\footnote{ ambuj\_math@rediffmail.com } Umesh Kumar Sharma$^{3}$\footnote{ sharma.umesh@gla.ac.in} }
\address{$^{1,2,3}$ Department of Mathematics, Institute of Applied Sciences and Humanities, GLA University\\
	Mathura-281406, Uttar Pradesh, India.}

\begin{abstract}
	\begin{center}
		{\bf Abstract}\\
	\end{center}
In this paper, we have examined the R$\acute{e}$nyi holographic dark energy (RHDE) model in the framework of an isotropic and spatially homogeneous flat FLRW (Friedmann- Lema$\hat i$tre-Robertson-Walker) Universe by considering different values of parameter $\delta$, where the infrared cut-off is taken care by the Hubble horizon. We examined the RHDE model through the analysis of the growth rate of perturbations and the statefinder hierarchy. The evolutionary trajectories of the statefinder hierarchy $S_3^1$, $S_3^2$ $S_4^1$, $S_4^2$ versus redshift $z$, shows satisfactory behaviour throughout the Universe evolution. One of the favourable appliance for exploring the dark energy models is the  CND (composite null diagnostic) $\{ S_3^1 - \epsilon\}$ and $\{ S_4^1 - \epsilon\}$, where the evolutionary trajectories of the  $\{ S_3^1 - \epsilon\}$  and $\{ S_4^1 - \epsilon\}$ pair show remarkable characteristics and the departure from $\Lambda$CDM could be very much assessed.\\
\end{abstract}
\maketitle
\section{Introduction}

Presently cosmologists are facing the problem to understand the reason behind the cosmic acceleration \cite{ref1,ref2}. This can be explained by one of the methods known as the concept of dark energy (DE).  But the nature of DE is not known yet. Various experiments like LSST  \cite{ref3}, DES \cite{ref4},  WFIRST \cite{ref5} and DESI \cite{ref6} will survey the Universe to understand the nature of DE.  Adjoint to these surveys, the accelerated expansion of the universe are supported by various observations like CMBR anisotropies,  BAO, SNeIa and LSS formation and WL, which is also consistent with the current standard cosmological model  $\Lambda$CDM, where $\Lambda$ is a constant component of cosmological fluid with the equation of state EoS  $\omega = -1$. Evolution of the cosmological constant is characterized as DE. Since we can only notice the impact of DE on the Hubble flow measuring observable components like radiation and matter. Therefore, we can not evaluate DE directly. Since the DE and gravity are associated in the standard model, we can compute the distribution of matter in the universe by $\omega = p(\rho)$, where $(\rho)$ is the matter-energy density and $p$ is the pressure. For simplicity, we take $\omega = p/\rho$.\\

Currently, discussion is around on the authenticity of $\Lambda$CDM model, which resulting disagreement between Plank \cite{ref7} and other cosmological measurements like strong lensing time delays (H0LiCOW),  megamasers, Cefeids (SH0ES), tip of the red giant branch (TRGB) and Oxygenrich Miras and surface brightness fluctuations \cite{ref8}. These disagreements encourages one to examine other options to the concordance model. To imitate $\Lambda$ according to the cosmological observations at the present time, dynamical dark energy models are most interesting approach.  Various approaches are extended theories of gravity \cite{ref9}, Bayesian reconstruction of a time-dependent EoS \cite{ref10}, dark energy parametrisations \cite{ref11,ref12,ref13,ref14,ref15}, $\omega(z)$ reconstructions \cite{ref16}, modify gravity \cite{ref17,ref18},
non-parametric reconstructions of $\omega(z)$ \cite{ref19,ref20},
quintessence scenarios \cite{ref21,ref22} and dynamical $\omega_x$  from $f(R)$ models \cite{ref23,ref24}. These models helps us in understanding the effects of DE. To describe the accelerated expansion of the cosmos motivated by holograpic principle \cite{ref25,ref26,ref27,ref28}, M. Li suggested Holographic dark energy (HDE) where IR cutoff was taken care by future event horizon \cite{ref29}. After that Agegrapic dark enrgy (ADE) model was suggested by Cai by taking length measure as the age of the Universe \cite{ref30}. By considering conformal time as time scale, Wei and Cai  suggested the New agegraphic dark energy (NADE) model  \cite{ref31}. Gao et al. \cite{ref32} suggested the Ricci dark energy model by replacing future event horizon with ricci scalar curvature.
For investigation of the cosmological and gravitational  and  incidences recently, different entropies \cite{ref33,ref34,ref35,ref36} has been used to find new form of DE models such as Tsallis holographic dark energy (THDE) model  \cite{ref37}, Tsallis agegraphic dark energy (TADE) model  \cite{ref38}, R$\acute{e}$nyi holograpic dark energy (RHDE) model \cite{ref39} and and Sharma-Mittal holograpic dark energy (SMHDE) model  \cite{ref40}. Researchers had used these newly proposed dark energy models in different scenarios\cite{ref41,ref42,ref43,ref44,ref45,ref46,ref47,ref48,ref49,ref50,ref51,ref52,ref53}.\\

Therefore, there is an absolute need for the diagnostic tools which can discriminate these various form of dark energy models. Keeping this in mind, different diagnostic tools are proposed to discriminate among various DE models. In \cite{ref54}, authors proposed the growth rate of linear perturbations and statefinder hierarchy, as null diagnostics to differentiate among different dark energy models from $\Lambda$CDM model. The statefinder hierarchy is a geometrical diagnostic which involves higher-order differential coefficients of scale factor $a(t)$, and also model-independent \cite{ref55}. Statefinder hierarchy has been used to discrimination among BMG (Bimetric Massive Gravity) theory, DGP models and  MGG (Modified Galileon Gravity) \cite{ref56}.  To differentiate the HDE models and breaking the degeneracy, the statefinder hierarchy has been investigated in \cite{ref57}.
To discriminate among purely kinetic $k$-essence,  modified Chaplygin gas,  superfluid Chaplygin gas, generalized Chaplygin gas and $\Lambda$CDM model and for analysing the deviation from the $\Lambda$CDM model, the statefinder hierarchy and growth rate of perturbation has been used \cite{ref58,ref59}. The statefinder hierarchy and the growth rate of perturbations are used in \cite{ref60,ref61,ref62,ref63,ref64,ref65,ref66,ref67,ref68,ref69,ref70,ref71,ref72}. Using statefinder hierarchy in the non-flat Universe considering IR cutoff as apparent horizon has been investigated by one of the authors of present work for the THDE models \cite{ref73}.\\

In this work, we have explored the newly proposed  R$\acute{e}$nyi  Holographic Dark Energy (RHDE) model through the diagnostic tools described above in the flat FRW Universe by taking the Hubble horizon as an infrared cutoff, which has not been explored earlier. Also, we have examined the deviation of the RHDE model from $\Lambda$CDM using these diagnostic tools. The paper is structured as follows; In Sect. II, we in brief visit the   R$\acute{e}$nyi holographic dark energy. Sect. III is dedicated to discussing the flat FRW cosmological model.  Section IV is divided into two subsections A and B, for the methods of the statefinder hierarchy diagnostic and growth rate of perturbations analysis. Finally, in the last section, we have given inferences.

\section{R$\acute{E}$nyi Holographic Dark Energy Model}

The form of the Bekenstein entropy of a system is $S = \frac{A}{4}$ , where $A = 4 \pi L^{2}$ and L is the IR cut-off. Another modified form of the R$\acute{e}$nyi entropy \cite{ref75} is given as: 

\begin{eqnarray}
\label{eq1}
S=\frac{1}{\delta}\log  \left(\frac{\delta}{4} A + 1\right ) = S=\frac{1}{\delta}\log  \left(\pi\delta L^{2}+ 1\right ),
\end{eqnarray} 
R$\acute{e}$nyi HDE density, by considering the assumption $ \rho _d\  dV \  \propto \   T dS$, takes the following form:

\begin{eqnarray}
\label{eq2}
\rho _D=\frac{3 c^2}{8 \pi L^{2}} (\pi\delta L^{2}+ 1)^{-1},
\end{eqnarray} 
By taking Hubble horizon as an IR cut-off $L = \dfrac{1}{H}$, we obtained:
\begin{eqnarray}
\label{eq3}
\rho _D=\frac{3 c^2 H^2}{8 \pi  \left(\frac{\pi  \delta }{H^2}+1\right)},
\end{eqnarray} 
where $ c^{2} $ is a numerical constant as usual.

\section{The cosmological model}
For  the flat  FRW  Universe, the metric is given  as :
\begin{eqnarray}
\label{eq4}
ds^{2} = -dt^{2}+a^{2}(t)\Big(dr^{2} + r^{2}d\Omega^{2}\Big).
\end{eqnarray}
In a flat FRW Universe, the  Friedmann first equation, involving DM and  RHDE is defined as :

\begin{eqnarray}
\label{eq5}
H^2 =\frac{1}{3} (8 \pi G) \left(\rho_D+\rho _m\right),
\end{eqnarray}

where $\rho_{D}$ and $\rho_{m}$  represent the energy density of  RHDE and matter, respectively.   The energy density parameter of  RHDE and  pressureless matter with the help of the fractional energy densities, can be defined  as \\
\begin{eqnarray}
\label{eq6}
\Omega_{m} = \frac{8\pi\rho_{m}G}{3H^{2}} , \hspace{1cm} \Omega_{D} = \frac{8\pi\rho_{D}G}{3H^{2}},
\end{eqnarray}

Now Eq. (\ref{eq5}) with help of  Eq. (\ref{eq6}) can be written as:
\begin{eqnarray}
\label{eq7}
1 = \Omega _D+\Omega _m
\end{eqnarray}
The  conservation law for matter and RHDE  are given as :
\begin{eqnarray}
\label{eq8}
\dot \rho_{m} + 3 H \rho_{m} = 0,
\end{eqnarray}
\begin{eqnarray}
\label{eq9}
\dot \rho_{D} + 3H (\rho_{D} + p_{D}) = 0.
\end{eqnarray}
in which $ \omega _D = p _D/\rho _D$ represents the RHDE EoS parameter. Now, using differential with time of Eq. (\ref{eq5}) in Eq. (\ref{eq8}), and Eq. (\ref{eq9}) combined the result with the Eq. (\ref{eq7}), we get\\

\begin{eqnarray}
\label{eq10}
\frac{\dot{H}}{H^2}=  \frac{3}{2} \left(\frac{\pi  \delta  \Omega _D}{\pi  \delta  \left(2 \Omega _D-1\right)+\left(\Omega _D-1\right) H(t)^2}-1\right)
\end{eqnarray}
By Eq. (\ref{eq10}), the deceleration parameter (DP) $q$ is found as\\

$q=-1 -\frac{\dot{H}}{H^2}$
\begin{eqnarray}
\label{eq11}
q =\frac{(1-2 \delta ) \Omega _D(t)+1}{2 \left(1-(2-\delta ) \Omega _D(t)\right)}
\end{eqnarray}
Also, by taking the derivative with respect to time of Eq. (\ref{eq3}),  we get
\begin{eqnarray}
\label{eq12}
\dot{\rho _D}=2 \rho _D \frac{\dot{H}}{H}    \left(\frac{\pi  \delta }{\pi  \delta +H^2}+1\right)
\end{eqnarray}

Now by using the Eqs. (\ref{eq12}) with  Eqs. (\ref{eq9}) and (\ref{eq10}), we gets expression for EoS parameter as:\\

\begin{eqnarray}
\label{eq13}
\omega _D=-\frac{\pi  \delta }{\pi  \delta  \left(2 \Omega _D-1\right)+H^2 \left(\Omega _D-1\right)}.
\end{eqnarray}
Also, taking the time differential  of the energy density parameter $\Omega _D $ with Eqs. (\ref{eq10}) and (\ref{eq12}), we find\\

\begin{eqnarray}
\Omega _D'=-\frac{3 \pi  c^2 \delta  H^2 \left(\Omega _D-1\right)}{\left(\pi  \delta +H^2\right) \left(\pi  \delta  \left(2 \Omega _D-1\right)+H^2 \left(\Omega _D-1\right)\right)}
\end{eqnarray}

where the dot is the derivative while considering time and prime lets us obtain the derivative concerning ln a.\\

\begin{figure}[htp]
	\begin{center}
		\includegraphics[width=9cm,height=9cm]{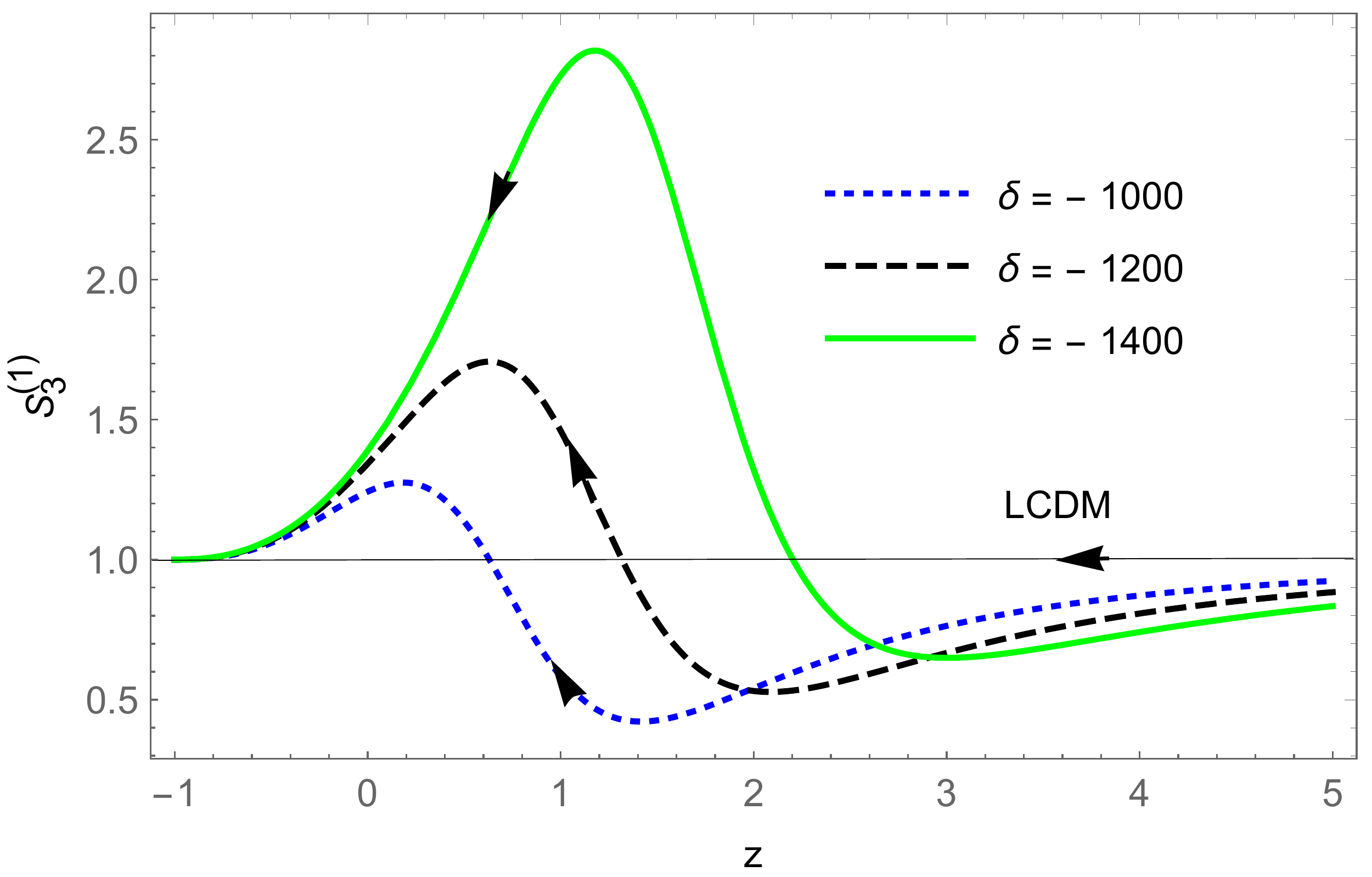}
		
		\caption{Graph of $S_3^{(1)}$  versus redshift z, for non- interacting RHDE with Hbbble radius as the IR cutoff. Here, $H(z=0)= 70$, $\Omega_{D}(z=0) = 0.73$ and different 
			values of $\delta$.}
		\label{fig:figure1}
	\end{center}
\end{figure}

\section{The methods of diagnostic} 

In this work, we used two diagnostic tools, statefinder hierarchy and growth rate of perturbations. We shall explore the RHDE model to discriminate from $\Lambda$CDM model with the help of these two diagnostic tools in this section.
\subsection{The Statefinder Hierarchy diagnostic}

\begin{figure}	
	\begin{center}
		\includegraphics[width=9cm,height=9cm]{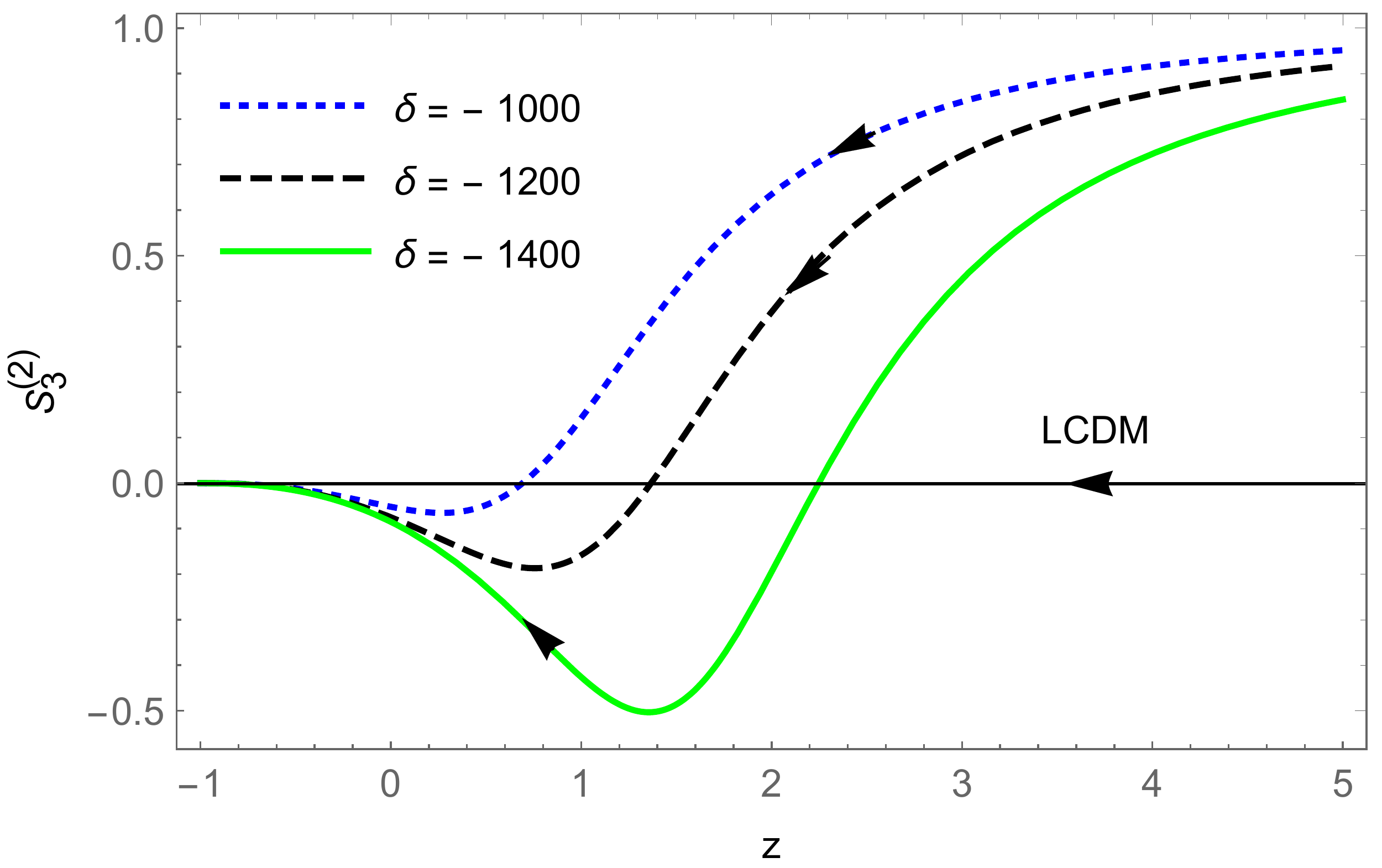}
		
		\caption {Graph of $S_3^{(2)}$  versus redshift z, for non- interacting RHDE with Hbbble radius as the IR cutoff. Here, $H(z=0)= 70$, $\Omega_{D}(z=0) = 0.73$ and different 
			values of $\delta$.}
	\end{center}
	\label{fig:figure2}
\end{figure}

\begin{figure}	
	\begin{center}
		\includegraphics[width=9cm,height=9cm]{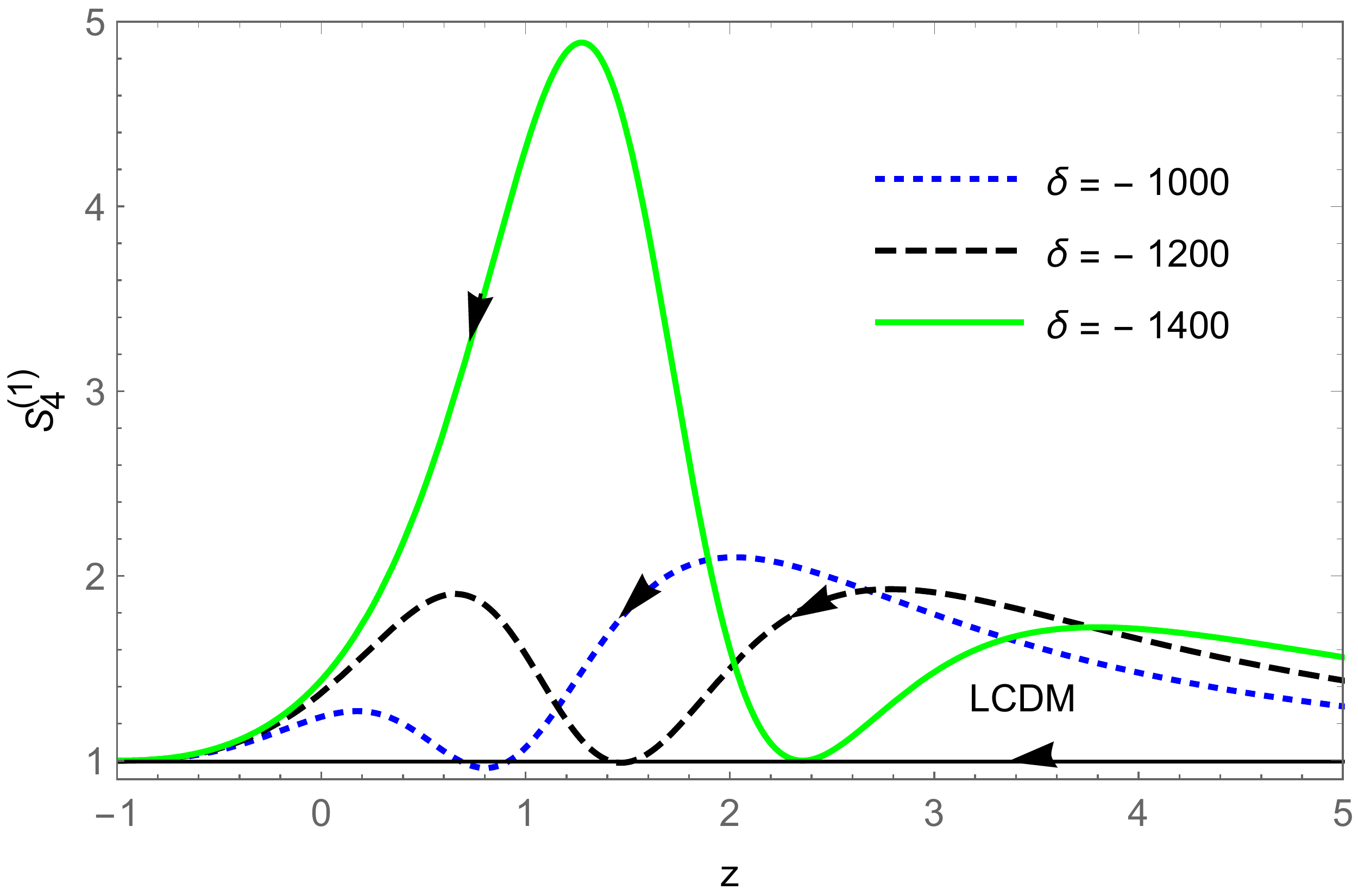}
		
		\caption {Graph of $S_4^{(1)}$  versus redshift z, for non- interacting RHDE with Hbbble radius as the IR cutoff. Here, $H(z=0)= 70$, $\Omega_{D}(z=0) = 0.73$ and different 
			values of $\delta$.}
	\end{center}
	\label{fig:figure3}
\end{figure}

\begin{figure}	
	\begin{center}
		\includegraphics[width=9cm,height=9cm]{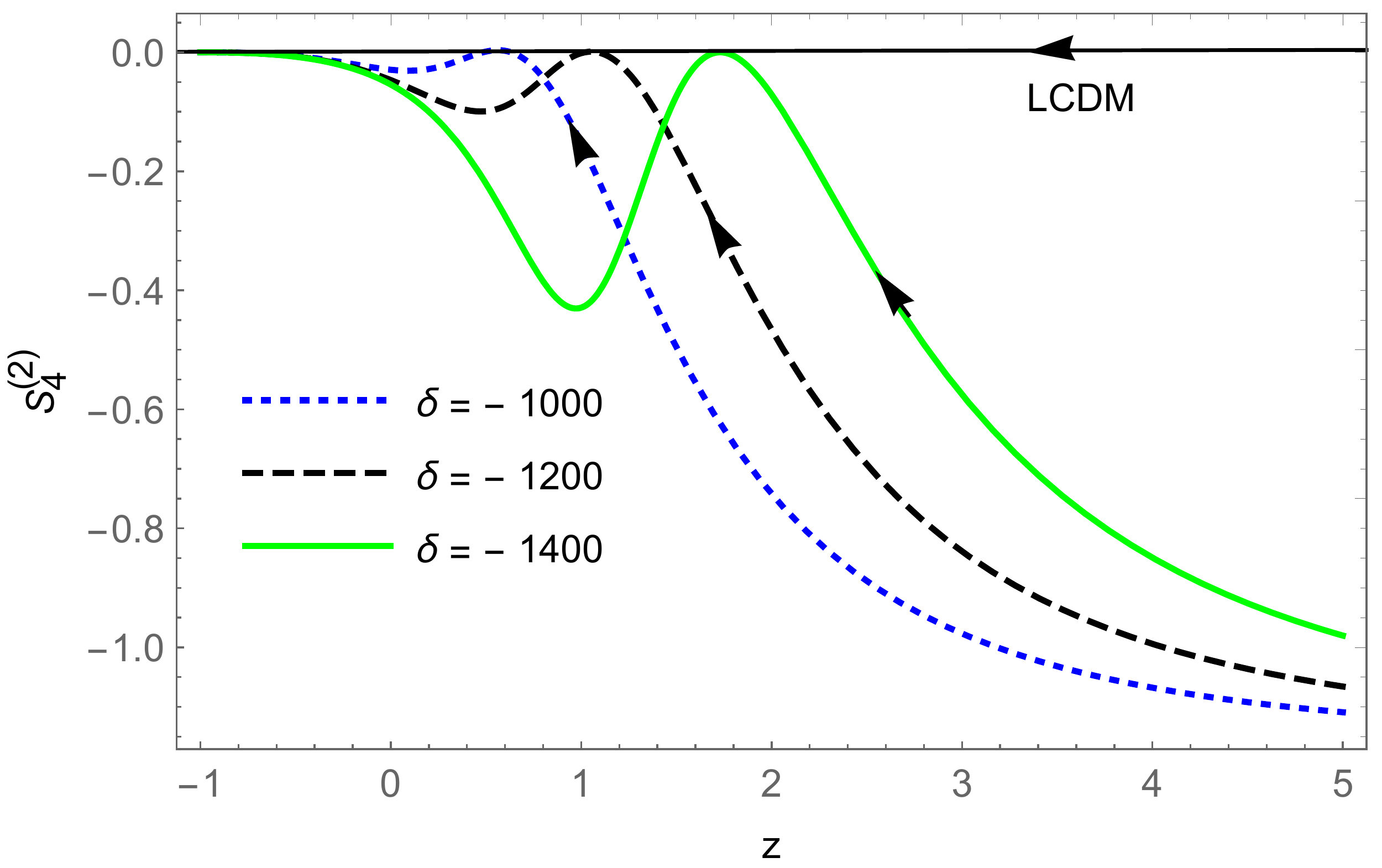}
		
		\caption {Graph of $S_4^{(2)}$  versus redshift z, for non- interacting RHDE with Hbbble radius as the IR cutoff. Here, $H(z=0)= 70$, $\Omega_{D}(z=0) = 0.73$ and different 
			values of $\delta$.}
	\end{center}
	\label{fig:figure4}
\end{figure}

Here, statefinder hierarchy diagnostic will be reviewed for the RHDE model will be described. The Taylor expansion of the scale factor $\frac{a(t)}{a_0}=\frac{1}{z+1}$, around the present epoch ${t_0}$ is given as:
\begin{eqnarray}
\label{eq15}
\frac{a (t)}{a_0}=\sum _{n=1}^{\infty } \frac{ A_n(t_0)}{n!}\left[H_0 \left(t-t_0\right)\right]{}^n
\end{eqnarray}

Where $A_n=\frac{a^n}{a H^n}$, ${a^n}$ is the $n^{th}$ derivative of the scale factor a verses cosmic time t and n $\in$ N. The statefinder hierarchy $S_{n}$ is defined as follows \cite{ref75a}:
\begin{eqnarray}
\label{eq16}
S_2=A_2+\frac{3 \Omega _m}{2}, S_3=A_3\quad and\quad   S_4=A_4+\frac{9 \Omega _m}{2},
\end{eqnarray}

Aforementioned gives the diagnostics for the model ($\Lambda$CDM) with $n \geq 3$, i.e., $S_{n}|\Lambda$CDM = 1. Hence by the use of $\Omega _m=\frac{2 (q+1)}{3}$ the statefinder hierarchy $S_3^{(1)}$, $S_4^{(1)}$ can be written as:

\begin{eqnarray}
\label{eq17}
S_3^{(1)}=A_3, \quad and \quad S_4^{(1)}=A_4+3 (q+1),
\end{eqnarray}

For $\Lambda$CDM model, $S_{n}^{(1)}$ = 1. In  \cite{ref72} it gives a path for construction of second Statefinder $S_3^{(1)}=S_3$ namely
\begin{eqnarray}
\label{eq18}
S_3^{(2)}=\frac{S_3^{(1)}-1}{3 \left(q-\frac{1}{2}\right)}
\end{eqnarray}

In concordance cosmology $S_3^{(1)} = 1$ while $S_3^{(2) = 0}$. Hence, $\left\{S_3^{(1)},S_3^{(2)}\right\}=\{1,0\}$ gives a model independent
means for forming a distinction between the dark energy models
from the cosmological constant \cite{ref72}. Eq. (\ref{eq18}) gives the second member of the Statefinder hierarchy

\begin{eqnarray}
\label{eq19}
S_n^{(2)}=\frac{S_n^{(1)}-1}{\alpha \left(q-\frac{1}{2}\right)},
\end{eqnarray}
where $\alpha$ is an arbitrary constant. In concordance cosmology $S_n^{(2)}= 0$ and
\begin{eqnarray}
\label{eq20}
\left\{S_n^{(1)},S_n^{(2)}\right\}=\{1,0\},
\end{eqnarray}
Some of degeneracies in $S_n^{(1)}$ can be removed by using the second statefinder $S_n^{(2)}$. For the dark energy model, we have
\begin{eqnarray}
\label{eq21}
S_3^{(1)}=\frac{1}{2} \left(9 \omega _D\right) \left(\omega _D+1\right) \Omega _D+1
\end{eqnarray}
\begin{eqnarray}
\label{eq22}
S_3^{(2)}=\omega _D+1
\end{eqnarray}

$S_4^{(1)} =-\frac{1}{4} \left(27 \omega _D^2\right) \left(\omega _D+1\right) \Omega ^2{}_D -  $\\
\begin{eqnarray}
\label{eq23}
\frac{1}{2} \left(27 \omega _D\right) \left(\omega _D+1\right) \left(\omega _D+\frac{7}{6}\right) \Omega _D+1
\end{eqnarray}
\begin{eqnarray}
\label{eq38}
S_4^{(2)}=-\frac{1}{2} \omega _D \left(\omega _D+1\right) \Omega _D-\left(\omega _D+1\right) \left(\omega _D+\frac{7}{6}\right)
\end{eqnarray}where $S_4^{(2)}=\frac{S_4^{(1)}-1}{9 \left(q-\frac{1}{2}\right)}$ and $q-\frac{1}{2}=\frac{1}{2} \left(3 \omega _D\right) \Omega _D$. As we demonstrate in figures 1, 2, 3, 4 the Statefinder hierarchy $\left\{S_n^{(1)},S_n^{(2)}\right\}$ 
give us a nice way to differenciating  dynamical dark energy models from $\Lambda$CDM model.\\ 

Fig. 1, shows the evolutionary trajectories of $S_3^{(1)}(z)$ for the RHDE model by considering different values of $\delta$. The separation of curvilinear shape is more distinct of the RHDE model in the region $0\leq z \leq 3$ for different values of $\delta$. It is observed that all the curves of $S_3^{(1)}(z)$ starts below the $\Lambda$CDM  line $S_3^{(1)} = 1$ and  monotonically increases by crossing the $\Lambda$CDM  line $S_3^{(1)}=1$ form convex vertices in the region $0\leq z \leq 3$ and then follow the close degeneration together into $\Lambda$CDM $S_3^{(1)}=1$, at low-redshift region.
The curves of $S_3^{(1)}(z)$ discriminate well from $\Lambda$CDM in the high-redshift region but highly degenerate in low-redshift region. This also shows that  different values of $\delta$  has quantitative impacts on the $S_3^{(1)}(z)$.\\

Fig. 2, shows the evolutionary trajectories of $S_3^{(2)}(z)$ for the RHDE model in the framework of an isotropic and spatially homogeneous flat FRW Universe by considering different values of parameter $\delta$, where the infrared cut-off is taken care by the Hubble horizon. We observe that 
the evolutionary trajectories of $S_3^{(2)}(z)$ are well differentiated from  $\Lambda$CDM line $S_3^{(2)}=0$, at high red-shift region. It starts descending monotonically from the high-redshift region and by crossing the $\Lambda$CDM line $S_3^{(2)}=0$, forms concave vertices   
and degenerate closely together with $\Lambda$CDM  $S_3^{(2)}=0$, at low-redshift region. These results shows that  different values of $\delta$ has quantitative impacts on the $S_3^{(2)}(z)$.\\

In Fig. 3, we give  the graph for $S_4^{(1)}$ evolution versus $z$ i.e. redshift for the RHDE model by considering different values of $\delta$. We notice that evolutionary trajectories of $S_4^{(1)}$ evolves above the $\Lambda$CDM line $S_4^{(1)}=1$, at high-redshift region and monotonically increases. Forming first concave vertex by touching the $\Lambda$CDM line and again it increases monotonically and forms convex vertices. While the evolutionary trajectories of $S_4^{(1)}$ for $\delta$=-1200, crosses the $\Lambda$CDM line. Finally these curves degenerate closely together with $\Lambda$CDM  $S_4^{(1)} 1$, at low-redshift region. We notice only quantitative impact on the $S_3^{(2)}(z)$ by varying $\delta$.\\ 

In Fig. 4, we give the graph for  $S_4^{(2)}$(z) evolution versus $z$ i.e. redshift for the RHDE model by considering different values of $\delta$. It is well-differentiated and evolves below the  $\Lambda$CDM line $S_4^{(2)}=0$ at the high red-shift region, and increases monotonically and form convex vertices at $\Lambda$CDM line for all values of  $\delta$. Then it decreases monotonically, again by making concave vertices finally degenerate closely together with $\Lambda$CDM line at the low red-shift region. These evolutionary trajectories shows only quantitative impact on the $S_4^{(2)}(z)$ by varying $\delta$.\\

From figures 1-4, we observe that all the curves superposing at $\Lambda$CDM line in the low-redshift region and all the curves separate well in the high-redshift region. So there are two shortcomings in the figures of $S_3^{(1)}(z)$ and  $S_4^{(1)}(z)$.
Hence the single diagnostic of geometry is not sufficient. It will be better to combine with the fractional growth parameter, as CND for getting more clear discrimination.\\

\begin{table}
	\caption{\small The present values of statefinders and fractional growth parameter,
		$S_{30}^{(1)}$, $S_{40}^{(1)}$ , and $ \epsilon_0$, and the differences of them, $ \bigtriangleup S_{30}^{(1)}$,  $ \bigtriangleup S_{40}^{(1)}$, and	$ \bigtriangleup  \epsilon_0$}
	\begin{center}
		\begin{tabular}{ |c |c |c|c|}
			\hline
			$\delta$	&	$- 1000$ &  $- 1200$ & $- 1400 $  \\
			
			\hline				
			$S_{30}^{(1)}$ &   $1.24276$ &  $1.34219$ &  $ 1.38857$	 \\
			
			\hline				
			$S_{40}^{(1)}$	&   $1.23685$ &  $1.36591$ &  $1.43563$	 \\
			
			\hline				
			$ \epsilon_0$	&   $0.995679$ &  $0.996293$ &  $0.996101$	 \\
			\hline				
			$ \bigtriangleup S_{30}^{(1)}$	& \multicolumn{3}{c|}{0.145811 }  \\	
			\hline				
			$ \bigtriangleup S_{40}^{(1)}$	& \multicolumn{3}{c|}{0.198776 }\\			
			\hline				
			$ \bigtriangleup  \epsilon_0$	& \multicolumn{3}{c|}{ 0.000421673 }  \\	
			\hline

		\end{tabular}

	\end{center}	
\end{table}	
\begin{figure}	
	\begin{center}
		\includegraphics[width=9cm,height=9cm]{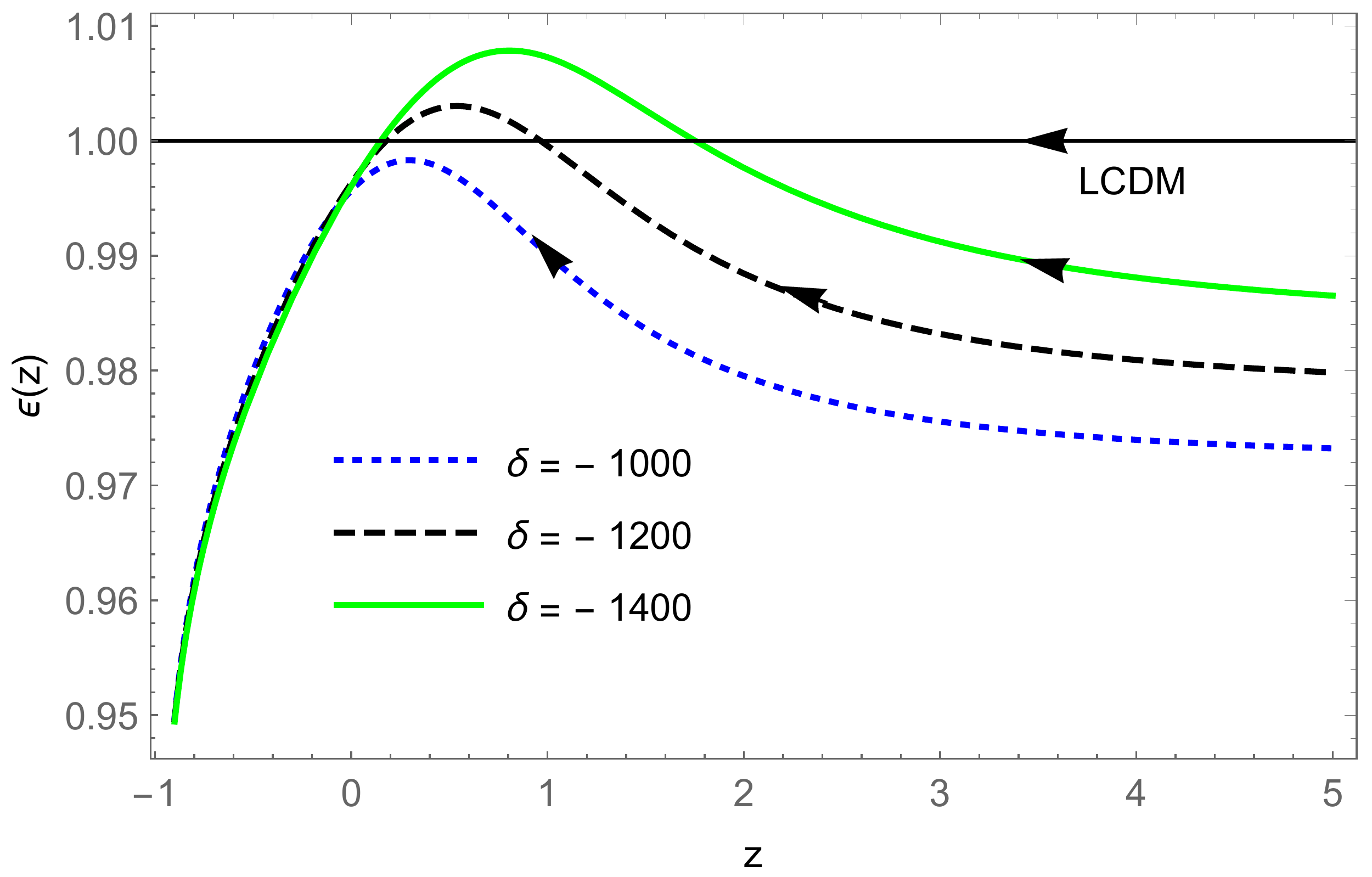}
		
		\caption {Graph of  $\epsilon(z)$ versus redshift z, for non- interacting RHDE with Hbbble radius as the IR cutoff. Here, $H(z=0)=70$, $\Omega_{D}(z=0)=0.73$ and different 
			values of $\delta$.}
	\end{center}
	\label{fig:figure5}
\end{figure}

\begin{figure}	
	\begin{center}
		\includegraphics[width=9cm,height=9cm]{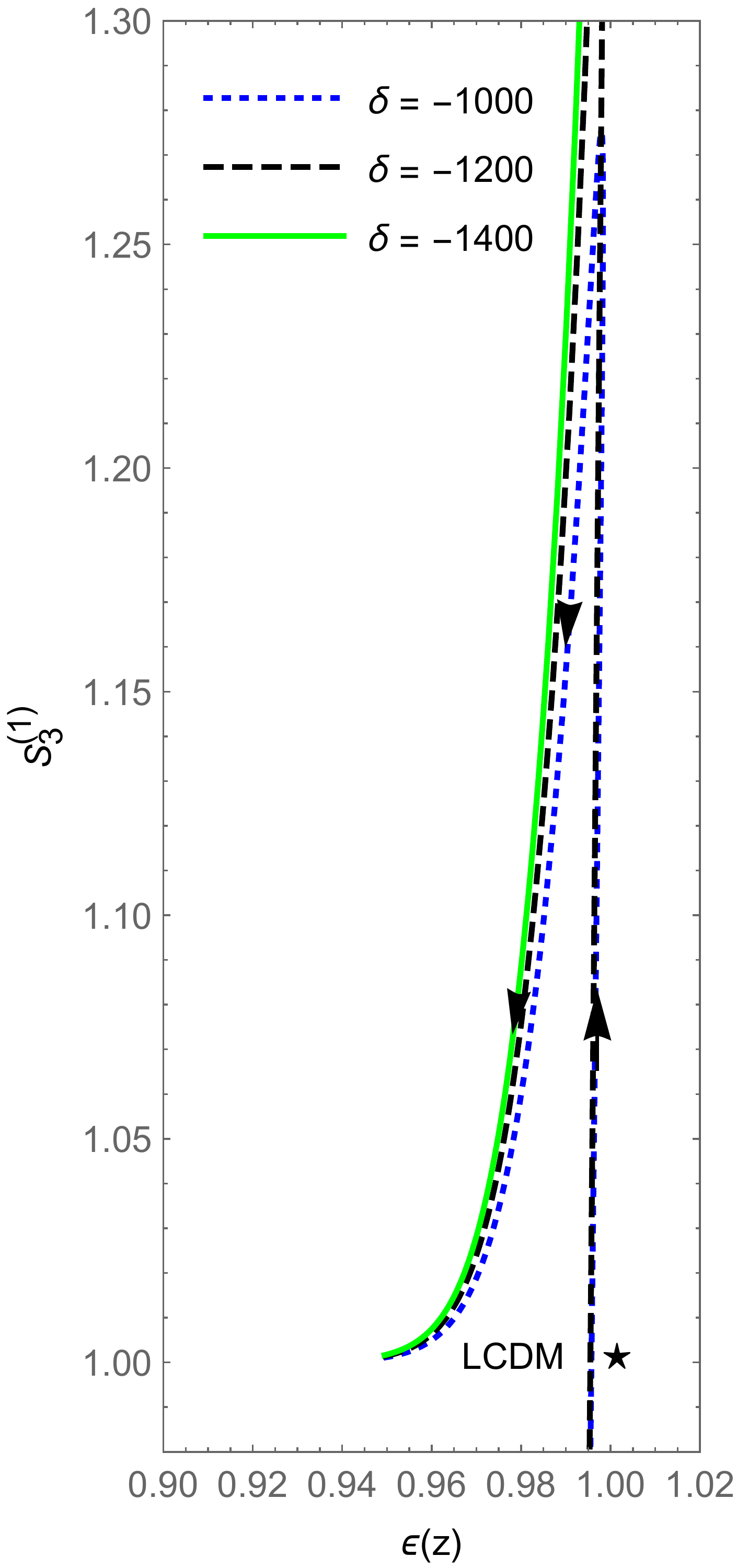}
		
		\caption { Graph of  $S_3^{(1)}$  versus $\epsilon(z)$, for non- interacting RHDE with Hbbble radius as the IR cutoff. Here, $H(z=0)=67$, $\Omega_{m}(z=0)=0.26$, $R = 10000$ and different 
			values of  $\delta$ (upper panel) and  $H(z=0)=67$, $\Omega_{m}(z=0) 0.26$,  $\delta=-600$ and different 
			values of R (below panel).}
	\end{center}
	\label{fig:figure6}
\end{figure}

\begin{figure}	
	\begin{center}
		\includegraphics[width=9cm,height=9cm]{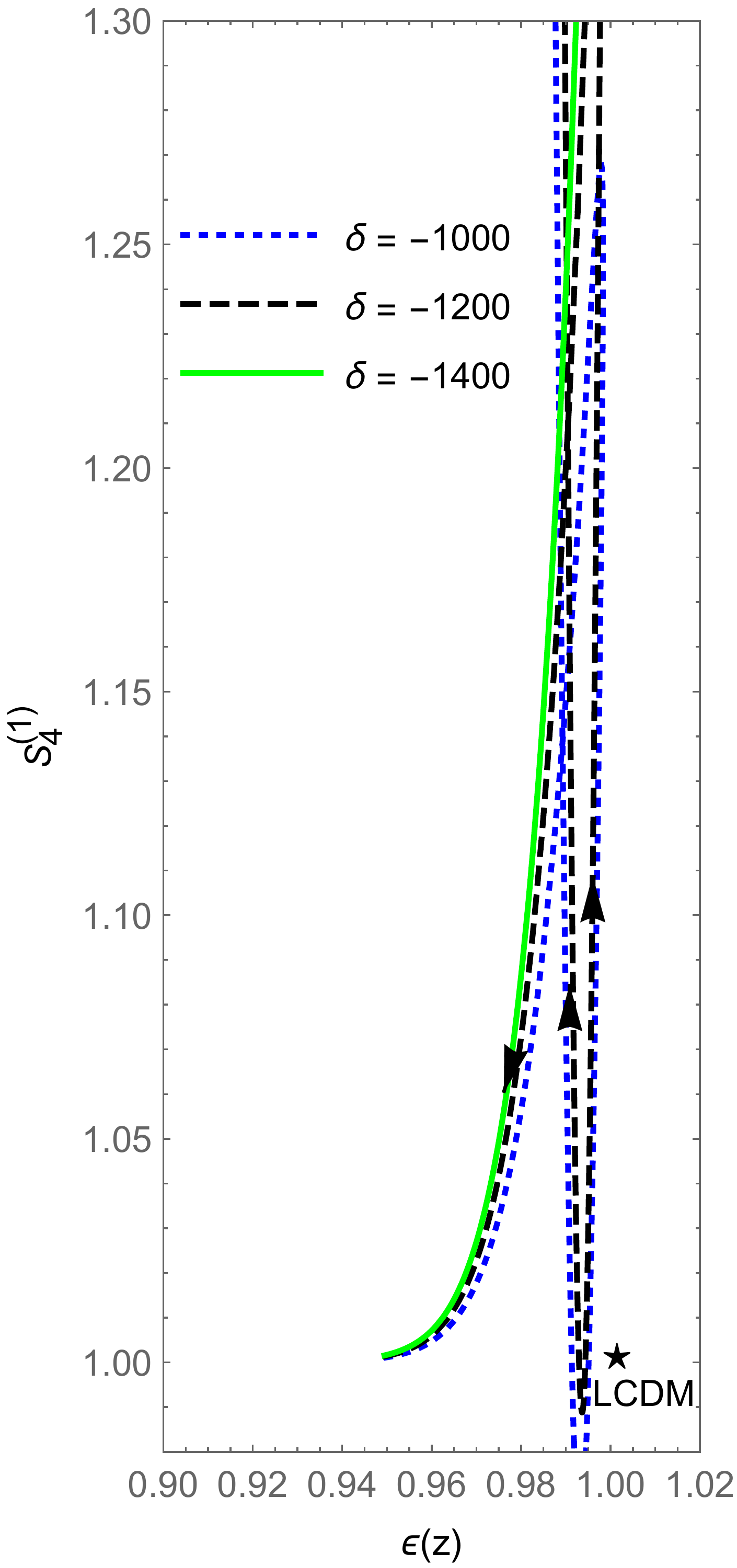}
		
		\caption {  Graph of  $S_4^{(1)}$  versus $\epsilon(z)$, for non- interacting RHDE with Hbbble radius as the IR cutoff. Here, $H(z=0)= 70$, $\Omega_{D}(z=0) = 0.73$ and different 
			values of $\delta$.}
	\end{center}
	\label{fig:figure6}
\end{figure}

\subsection{Growth rate of perturbations}

The fractional growth parameter $ \epsilon(z)$  \cite{ref76,ref77} is determined as
\begin{eqnarray}
\label{eq25}
\epsilon(z) =\frac{f (z)}{ f_{\text{$\Lambda $CDM}}(z)}
\end{eqnarray}
Here $f (z)=\frac{d \log \delta }{d \log a }$  is the growth rate of structure. Here, $\delta =\frac{\delta  \rho _m}{\rho _m}$, with $\delta  \rho _m$  and  $\rho _m$ being the the density perturbation and energy density of matter (including CDM and baryons), respectively. If the perturbation is in the linear fashion and without any  interaction between DM and DE, then  we can say that the  equation of perturbation at late times can be:

\begin{eqnarray}
\label{eq26}
\ddot{\delta }+ 2 \dot{\delta } H= 4 \pi  \delta  G \rho _m
\end{eqnarray}

Here, Newton's gravitational constant is represented by $G$. So, the approx  growth rate of linear density perturbation can be reflected by \cite{ref78}:
\begin{eqnarray}
\label{eq27}
f (z)\simeq \Omega _m (z)^{\gamma }
\end{eqnarray}

$\gamma  (z)=\frac{\left(3 \left(1-\omega _D\right) \left(1-\frac{3 \omega _D}{2}\right)\right) \left(1- \Omega _m(z)\right)}{125 \left(1-\frac{6 \omega _D}{5}\right){}^3}$ +
\begin{eqnarray}
\label{eq28}
\frac{3}{5-\frac{\omega _D}{1-\omega _D}}
\end{eqnarray}

where $ \Omega _m(z)=\frac{ \rho _m(z)}{3 H (z)^2 M_p^2}$, the fractional density of matter, $\Omega$ is constant 
or varies slowly with time. $\epsilon(z)$ = 1 and  $\gamma \simeq 0.55$ are the values for the $\Lambda$CDM model \cite{ref78,ref79}. For other models $\epsilon(z)$ exhibits differences from $\Lambda$CDM which would be the possible reason for its use as a diagnostic. By applying the composite null diagnostic ${CND} \equiv \left\{S_{n,}\epsilon \right\}$ where
$ \left\{S_{n,}\epsilon \right\} = \{1,1\}$ for $\Lambda$CDM,  we can make use of both matter perturbational as well as geometrical information of cosmic evolution. While, we can analyze and present only one-side information of cosmic evolution by using one single diagnostic tool. \\

For the diagnose of diverse theoretical DE models,  
the evolution of the fractional growth parameter $\epsilon (z)$ is analysed.  The evolutionary trajectories of  $\epsilon (z)$ versus redshift $z$ for a spatially homogeneous and an isotropic flat FRW Universe of RHDE model by considering different values of $\delta$ are plotted in Fig. 5. It is observed that the evolutionary trajectories of $\epsilon (z)$ evolves below the $\Lambda$CDM line  $\epsilon (z) = 1$ and the curves separate well at the high red-shift region. The curves for  $\delta$ = - 1200,  $\delta$ = - 1400 are monotonically increasing and form convex vertices by crossing the  $\Lambda$CDM line from past to present. The evolutionary trajectories of  $\epsilon (z)$ for  $\delta$ = - 1000, behaves in the same way but it does not cross the  $\Lambda$CDM line. Presently, these curves are degenerated together and decrease monotonically for the low red-shift region.\\


Fig. 6 shows the evolutionary trajectories of $\left\{S_3^{(1)},\epsilon \right\}$ of RHDE model by considering different values of $\delta$. Where star symbol denotes the $\Lambda$CDM model $\left\{S_3^{(1)}= 1,\epsilon = 1\right\}$. From figure we observe  that, curve evolves near  $\Lambda$CDM model and monotonically increases and forms convex vertices and finally all these curves degenerated together to the line $S_3^{(1)}= 1$ at low red-shift region.\\

Fig. 7 is the the evolutionary trajectories of the CND pair $\left\{S_4^{(1)},\epsilon \right\}$  for the RHDE model by considering different values of $\delta$ (upper panel) and R (below panel).
The evolutionary trajectories of  $\left\{S_4^{(1)},\epsilon \right\}$ shows similar characteristic as the curves of $\left\{S_3^{(1)},\epsilon \right\}$. These results shows that adopting different values of $\delta$ has quantitative impacts and the deviation from $\Lambda$CDM can be seen in this figure.\\ 

The evolutionary trajectories of the CND pair $\left\{S_4^{(1)},\epsilon \right\}$  for the RHDE model by considering different values of $\delta$ are depicted by Fig. 7. The behaviour of  $\left\{S_4^{(1)},\epsilon \right\}$ are similar to the  $\left\{S_3^{(1)},\epsilon \right\}$.
In the Research work of cosmology, the present values of the parameters keeps important. In this direction we calculated the the present values
of parameters 
$S_{30}^{(1)}$, $S_{40}^{(1)}$ , and $ \epsilon_0$ of  the RHDE model by considering different values of $\delta$ and also the differences of them, for each case
$ \bigtriangleup S_{30}^{(1)}$ =  $S_{30}^{(1)}$(max)- $S_{30}^{(1)}$(min), $ \bigtriangleup S_{40}^{(1)}$ =  $S_{40}^{(1)}$(max)- $S_{40}^{(1)}$(min), and 	$ \bigtriangleup  \epsilon_0$= $ \epsilon_0$(max) - $ \epsilon_0$(min), which is given in Table 1. We observe  that $ \bigtriangleup S_{40}^{(1)}= 0.198776 >  \bigtriangleup S_{30}^{(1)} = 0.145811$.
We can see  $ \bigtriangleup S_{40}^{(1)} >  \bigtriangleup S_{30}^{(1)}$, which means that the fourth-order derivative of the scale factor in comparison to third-order derivative, intensifies the degeneracy of present values.  Therefore $S_{40}^{(1)}$ in comparison to
$S_{30}^{(1)} $ gives more variance for different values of $\delta$ among the cosmic evolution of RHDE type of DE. Which helps us to distinguish different theoretical models.\\

\section{Conclusions}
The paper uses the R$\acute{e}$nyi Holographic Dark Energy model in the framework of an isotropic and spatially homogeneous flat FRW Universe by considering different values of RHDE parameter $\delta$, where the infrared cut-off is taken care by the Hubble horizon. This can be summarized as \\
\begin{itemize}
	\item 
	In this paper, we examined the deviation of RHDE model from  $\Lambda$CDM with statefinder hierarchy supplemented by the growth rate of perturbations. 
	\item 
	The statefinder hierarchy $S_3^{(1)}$, $S_3^{(2)}$, $S_4^{(1)}$ and $S_4^{(2)}$,
	which contain the third and fourth derivatives of the scale factor, have been plotted versus red-shift z. The evolutionary trajectories of $S_3^{(1)}$ evolve below the $\Lambda$CDM line while
	$S_4^{(1)}$ evolves from above the $\Lambda$CDM line. The separation of the curvilinear shape of both parameters is more distinct in the region $0\leq z \leq 3$ for different values of $\delta$. The evolutionary trajectories of $S_3^{(2)}$ evolves above the $\Lambda$CDM line and crosses the $\Lambda$CDM line but $S_4^{(1)}$ evolves below the $\Lambda$CDM line and never crosses the $\Lambda$CDM line. All parameters degenerated closely together into $\Lambda$CDM line, at the low-redshift region.\\
	\item 
	We have also examines the growth rate of structure $\epsilon (z)$ by plotting it versus red-shift z along with the combination of statefinder hierarchy $S_n$. The evolutionary trajectories of $\epsilon (z)$ degenerated together and decrease monotonically for low red-shift region. The curves of $\epsilon (z)$ separate well at high red-shift region for different values of $\delta$. The evolutionary trajectories of  $\left\{S_3^{(1)},\epsilon \right\}$. and  $\left\{S_4^{(1)},\epsilon \right\}$, shows the same deviation from $\Lambda$CDM model of RHDE model for all values of $\delta$.\\	 
	\item 	 	
	For alleviating the degeneracy existing in other statefinder parameters
	for DE models, comparison of the present-value differences of the parameters $S_{30}^{(1)}$, $S_{40}^{(1)}$ plays an important role. By using CND, we can discriminate RHDE model from  $\Lambda$CDM. Since $ \bigtriangleup S_{40}^{(1)} >  \bigtriangleup S_{30}^{(1)}$, hence we can say that the fourth-order hierarchy of statefinder is a better choice than the third-order hierarchy for the RHDE model. Therefore, the above investigation concludes that the higher-order statefinder hierarchy, with the growth rate of perturbations, can differentiate the RHDE model from the $\Lambda$CDM model and also from itself with different parameter values.\\
	
	We hope that in future high precision observations, for example, SNAP-type investigation can be equipped for deciding the cosmological parameters exactly and consequently identify the correct cosmological model and closer to understand the properties of the RHDE model.
\end{itemize}
\section*{Acknowledgments}
The authors are thankful for the help and support given by  GLA University, Mathura, India, in this research work.


\begin{thebibliography}{99}
	
	
	\bibitem {ref1}
 A.~G.~Riess {\it et al.} [Supernova Search Team],
``Observational evidence from supernovae for an accelerating universe and a cosmological constant,''
Astron.\ J.\  {\bf 116} (1998) 1009
doi:10.1086/300499
	\bibitem {ref2}
	  S.~Perlmutter {\it et al.} [Supernova Cosmology Project Collaboration],
	 ``Measurements of $\Omega$ and $\Lambda$ from 42 high redshift supernovae,''
	 Astrophys.\ J.\  {\bf 517} (1999) 565
	 doi:10.1086/307221
		\bibitem {ref3}
	https://www.lsst.org/
		\bibitem {ref4}
	https://www.darkenergysurvey.org/
		\bibitem {ref5}
	https://wfirst.gsfc.nasa.gov/
	\bibitem {ref6}
	http://desi.lbl.gov/

	\bibitem {ref7}
	 N. Aghanim et al. [Planck Collaboration], arXiv:1807.06209 [astro-ph.CO].
	\bibitem {ref8}
	L. Verde, T. Treu and A. G. Riess, arXiv:1907.10625 [astro-ph.CO].
\bibitem {ref9}
  S.~Capozziello, R.~D'Agostino and O.~Luongo,
 ``Extended Gravity Cosmography,''
 Int.\ J.\ Mod.\ Phys.\ D {\bf 28} (2019) no.10,  1930016
 doi:10.1142/S0218271819300167
 
\bibitem {ref10}
  G.~B.~Zhao {\it et al.},
 ``Dynamical dark energy in light of the latest observations,''
 Nat.\ Astron.\  {\bf 1} (2017) no.9,  627
 doi:10.1038/s41550-017-0216-z
 
 \bibitem {ref11}
   I.~Sendra and R.~Lazkoz,
  ``SN and BAO constraints on (new) polynomial dark energy parametrizations: current results and forecasts,''
  Mon.\ Not.\ Roy.\ Astron.\ Soc.\  {\bf 422} (2012) 776
  doi:10.1111/j.1365-2966.2012.20661.x
 \bibitem {ref12}
 G. B. Zhao, D. Bacon, R. Maartens, M. Santos and A. Raccanelli, arXiv:1501.03840 [astro-ph.CO].
 \bibitem {ref13}
 C.~Escamilla-Rivera,
``Status on bidimensional dark energy parameterizations using SNe Ia JLA and BAO datasets,''
Galaxies {\bf 4} (2016) no.3,  8
doi:10.3390/galaxies4030008
 \bibitem {ref14}
 M.~Rezaei, M.~Malekjani, S.~Basilakos, A.~Mehrabi and D.~F.~Mota,
``Constraints to Dark Energy Using PADE Parameterizations,''
Astrophys.\ J.\  {\bf 843} (2017) no.1,  65
doi:10.3847/1538-4357/aa7898
 \bibitem {ref15}
  C.~Escamilla-Rivera and S.~Capozziello,
 ``Unveiling cosmography from the dark energy equation of state,''
 Int.\ J.\ Mod.\ Phys.\ D {\bf 28} (2019) no.12,  1950154
 doi:10.1142/S0218271819501542
 \bibitem {ref16}
  J.~Alberto Vazquez, M.~Bridges, M.~P.~Hobson and A.~N.~Lasenby,
 ``Reconstruction of the Dark Energy equation of state,''
 JCAP {\bf 1209} (2012) 020
 doi:10.1088/1475-7516/2012/09/020
 \bibitem {ref17}
  L.~G.~Jaime, L.~Patiño and M.~Salgado,
 ``Note on the equation of state of geometric dark-energy in f(R) gravity,''
 Phys.\ Rev.\ D {\bf 89} (2014) no.8,  084010
 doi:10.1103/PhysRevD.89.084010
 \bibitem {ref18}
 R.~Lazkoz, M.~Ortiz-Baños and V.~Salzano,
``$f(R)$ gravity modifications: from the action to the data,''
Eur.\ Phys.\ J.\ C {\bf 78} (2018) no.3,  213
doi:10.1140/epjc/s10052-018-5711-6
 
 \bibitem {ref19}
 M.~Seikel, C.~Clarkson and M.~Smith,
``Reconstruction of dark energy and expansion dynamics using Gaussian processes,''
JCAP {\bf 1206} (2012) 036
doi:10.1088/1475-7516/2012/06/036
 \bibitem {ref20}
   A.~Montiel, R.~Lazkoz, I.~Sendra, C.~Escamilla-Rivera and V.~Salzano,
 ``Nonparametric reconstruction of the cosmic expansion with local regression smoothing and simulation extrapolation,''
 Phys.\ Rev.\ D {\bf 89} (2014) no.4,  043007
 doi:10.1103/PhysRevD.89.043007
 \bibitem {ref21}
  B.~Ratra and P.~J.~E.~Peebles,
 ``Cosmological Consequences of a Rolling Homogeneous Scalar Field,''
 Phys.\ Rev.\ D {\bf 37} (1988) 3406.
 doi:10.1103/PhysRevD.37.3406
 \bibitem {ref22}
 C.~Armendariz-Picon, V.~F.~Mukhanov and P.~J.~Steinhardt,
``A Dynamical solution to the problem of a small cosmological constant and late time cosmic acceleration,''
Phys.\ Rev.\ Lett.\  {\bf 85} (2000) 4438
doi:10.1103/PhysRevLett.85.4438
 \bibitem {ref23}
 S.~Capozziello,
``Curvature quintessence,''
Int.\ J.\ Mod.\ Phys.\ D {\bf 11} (2002) 483
doi:10.1142/S0218271802002025
 \bibitem {ref24}
 L.~G.~Jaime, M.~Jaber and C.~Escamilla-Rivera,
 ``New parametrized equation of state for dark energy surveys,''
 Phys.\ Rev.\ D {\bf 98} (2018) no.8,  083530
 doi:10.1103/PhysRevD.98.083530


	\bibitem{ref25} 
L.~Susskind,
``The World as a hologram,''
J.\ Math.\ Phys.\  {\bf 36} (1995) 6377
doi:10.1063/1.531249

\bibitem{ref26} 

P.~Horava and D.~Minic,
``Probable values of the cosmological constant in a holographic theory,''
Phys.\ Rev.\ Lett.\  {\bf 85} (2000) 1610
doi:10.1103/PhysRevLett.85.1610
\bibitem{ref27}
S.~D.~Thomas,
``Holography stabilizes the vacuum energy,''
Phys.\ Rev.\ Lett.\  {\bf 89} (2002) 081301.
doi:10.1103/PhysRevLett.89.081301
\bibitem{ref28}
S.~D.~H.~Hsu,
``Entropy bounds and dark energy,''
Phys.\ Lett.\ B {\bf 594} (2004) 13
doi:10.1016/j.physletb.2004.05.020

\bibitem{ref29}
M.~Li,
``A Model of holographic dark energy,''
Phys.\ Lett.\ B {\bf 603} (2004) 1
doi:10.1016/j.physletb.2004.10.014


	\bibitem{ref30}
R.~G.~Cai,
``A Dark Energy Model Characterized by the Age of the Universe,''
Phys.\ Lett.\ B {\bf 657} (2007) 228
doi:10.1016/j.physletb.2007.09.061

	\bibitem{ref31} 
H.~Wei and R.~G.~Cai,
``A New Model of Agegraphic Dark Energy,''
Phys.\ Lett.\ B {\bf 660} (2008) 113
doi:10.1016/j.physletb.2007.12.030

	\bibitem {ref32} 	
C.~Gao and F.~Wu {\it et al.}, 
``A Holographic Dark Energy Model from Ricci Scalar Curvature,''
Phys.\ Rev.\ D {\bf 79} (2009) 043511.
doi:10.1103/PhysRevD.79.043511
	\bibitem {ref33}
C.~Tsallis and L.~J.~L.~Cirto,
``Black hole thermodynamical entropy,''
Eur.\ Phys.\ J.\ C {\bf 73} (2013) 2487
doi:10.1140/epjc/s10052-013-2487-6

\bibitem {ref34}
C.~Tsallis,
``Possible Generalization of Boltzmann-Gibbs Statistics,''
J.\ Statist.\ Phys.\  {\bf 52} (1988) 479.
doi:10.1007/BF01016429
\bibitem {ref35}
A.R$\acute{e}$nyi , in Proceedings of the 4th Berkely Symposium on Mathematics, Statistics and Probability (University California
Press, Berkeley, CA, 1961) pp. 547–561.
\bibitem {ref36}
B.D. Sharma and D.P. Mittal, `` New non-additive measures of entropy for discrete probability distributions,''
J. Math. Sci.{\bf10} (1975) 28-40; B.D. Sharma, D.P. Mittal, J. Comb. Inf. Syst. Sci. 2 (1977) 122.

	\bibitem {ref37}
M.~Tavayef, A.~Sheykhi, K.~Bamba and H.~Moradpour,
``Tsallis Holographic Dark Energy,''
Phys.\ Lett.\ B {\bf 781} (2018) 195
doi:10.1016/j.physletb.2018.04.001

\bibitem {ref38} 
M.~Abdollahi Zadeh, A.~Sheykhi and H.~Moradpour,
``Tsallis Agegraphic Dark Energy Model,''
Mod.\ Phys.\ Lett.\ A {\bf 34} (2019) no.11,  1950086
doi:10.1142/S021773231950086X
  
	\bibitem {ref39}
	H.~Moradpour, S.~A.~Moosavi, I.~P.~Lobo, J.~P.~Morais Graça, A.~Jawad and I.~G.~Salako,
	``Thermodynamic approach to holographic dark energy and the Rényi entropy,''
	Eur.\ Phys.\ J.\ C {\bf 78} (2018) no.10,  829
	doi:10.1140/epjc/s10052-018-6309-8
	

	
	\bibitem {ref40}
	A.~Sayahian Jahromi, S.~A.~Moosavi, H.~Moradpour, J.~P.~Morais Graça, I.~P.~Lobo, I.~G.~Salako and A.~Jawad,
	``Generalized entropy formalism and a new holographic dark energy model,''
	Phys.\ Lett.\ B {\bf 780} (2018) 21
	doi:10.1016/j.physletb.2018.02.052
	
		\bibitem {ref41}
	A.~Jawad, K.~Bamba, M.~Younas, S.~Qummer and S.~Rani,
	``Tsallis, Rényi and Sharma-Mittal Holographic Dark Energy Models in Loop Quantum Cosmology,''
	Symmetry {\bf 10} (2018) no.11,  635.
	doi:10.3390/sym10110635  
	
	\bibitem{ref42} 
	S.~Nojiri, S.~D.~Odintsov and E.~N.~Saridakis,``Modified cosmology from extended entropy with varying exponent,''  {\it Eur.\ Phys.\ J.\ C}  {\bf79} (2019) 242.
	
	\bibitem{ref43}
	Q.~Huang, H.~Huang, J.~Chen, L.~Zhang and F.~Tu,
	``Stability analysis of a Tsallis holographic dark energy model,''
	{\it Class.\ Quant.\ Grav.}  {\bf 36}, no. 17, (2019) 175001. 
	S.~Ghaffari, H.~Moradpour, I.~P.~Lobo, J.~P.~Morais Graça and V.~B.~Bezerra,
	``Tsallis holographic dark energy in the Brans–Dicke cosmology,''
	Eur.\ Phys.\ J.\ C {\bf 78} (2018) no.9,  706
	doi:10.1140/epjc/s10052-018-6198-x
	\bibitem{ref44}
	E.~N.~Saridakis, K.~Bamba, R.~Myrzakulov and F.~K.~Anagnostopoulos,
	``Holographic dark energy through Tsallis entropy,''
	JCAP {\bf 1812} (2018) 012
	doi:10.1088/1475-7516/2018/12/012
	\bibitem{ref45}
	V.~C.~Dubey, S.~Srivastava, U.~K.~Sharma and A.~Pradhan,
	``Tsallis holographic dark energy in Bianchi-I Universe using hybrid expansion law with $k$-essence,''
	Pramana {\bf 93} (2019) no.5,  78.
	doi:10.1007/s12043-019-1843-y
	\bibitem{ref46}
	E.~Sadri,
	``Observational constraints on interacting Tsallis holographic dark energy model,''
	Eur.\ Phys.\ J.\ C {\bf 79} (2019) no.9,  762
	doi:10.1140/epjc/s10052-019-7263-9
	\bibitem{ref47}
	E.~M.~Barboza, Jr., R.~d.~C.~Nunes, E.~M.~C.~Abreu and J.~Ananias Neto,
	``Dark energy models through nonextensive Tsallis’ statistics,''
	Physica A {\bf 436} (2015) 301
	doi:10.1016/j.physa.2015.05.002
	\bibitem{ref48}
	Golanbari, T., Saaidi, K.,  Karimi, P. (2020). " R$\acute{e}$nyi entropy and the holographic dark energy in flat space time". arXiv preprint arXiv:2002.04097.
	\bibitem{ref49}
	Sharma, U. K.,  Dubey, V. C. (2020). "Interacting R$\acute{e}$nyi holographic dark energy with parametrization on the interaction term". arXiv preprint arXiv:2001.02368.
	\bibitem{ref50}
	S.~Ghaffari, A.~H.~Ziaie, V.~B.~Bezerra and H.~Moradpour,
	``Inflation in the  R$\acute{e}$nyi cosmology,''
	Mod.\ Phys.\ Lett.\ A {\bf 35} (2019) no.01,  1950341
	doi:10.1142/S0217732319503413
	
	\bibitem{ref51}
	V. C. Dubey,  et al. ``Tsallis holographic dark energy Models in axially symmetric space time." 	{\it Int. J. Geom. Methods Mod. Phys.} {\bf17} 1  (2020) 2050011.
	
	\bibitem{ref52} 	
	Y.~Aditya, S.~Mandal, P.~K.~Sahoo and D.~R.~K.~Reddy,
	``Observational constraint on interacting Tsallis holographic dark energy in logarithmic Brans–Dicke theory,''
	{\it Eur.\ Phys.\ J.\ C} {\bf 79}12, 1020 (2019). 
	\bibitem{ref53}	
	A.~Iqbal and A.~Jawad,
	``Tsallis, Renyi and Sharma–Mittal holographic dark energy models in DGP brane-world,''
	Phys.\ Dark Univ.\  {\bf 26} (2019) 100349.
	doi:10.1016/j.dark.2019.100349
	
	
	
	
		\bibitem{ref54}
	M.~Arabsalmani and V.~Sahni,
	``The Statefinder hierarchy: An extended null diagnostic for concordance cosmology,''
	Phys.\ Rev.\ D {\bf 83} (2011) 043501.
	doi:10.1103/PhysRevD.83.043501
	\bibitem{ref55}
	J.~F.~Zhang and J.~L.~Cui {\it et al.},
	``Diagnosing holographic dark energy models with statefinder hierarchy,''
	Eur.\ Phys.\ J.\ C {\bf 74} (2014) no.10,  3100.
	doi:10.1140/epjc/s10052-014-3100-3
	
	
	
	
	
	\bibitem{ref56}
 R.~Myrzakulov and M.~Shahalam,
``Statefinder hierarchy of bimetric and galileon models for concordance cosmology,''
JCAP {\bf 1310} (2013) 047
doi:10.1088/1475-7516/2013/10/047
	
	\bibitem{ref57}
	 J.~F.~Zhang, J.~L.~Cui and X.~Zhang,
	``Diagnosing holographic dark energy models with statefinder hierarchy,''
	Eur.\ Phys.\ J.\ C {\bf 74} (2014) no.10,  3100
	doi:10.1140/epjc/s10052-014-3100-3
	
	\bibitem{ref58}
J. L.Cui, et al.,  "A closer look at interacting dark energy with statefinder hierarchy and growth rate of structure". JCAP {\bf09}(2015) 024.
	
	\bibitem{ref59}
	  J.~Li, R.~Yang and B.~Chen,
	``Discriminating dark energy models by using the statefinder hierarchy and the growth rate of matter perturbations,''
	JCAP {\bf 1412} (2014) 043
	doi:10.1088/1475-7516/2014/12/043

	\bibitem{ref60}
	V.~Acquaviva and A.~Hajian {\it et al.},
	``Next Generation Redshift Surveys and the Origin of Cosmic Acceleration,''
	Phys.\ Rev.\ D {\bf 78} (2008) 043514.
	doi:10.1103/PhysRevD.78.043514
	\bibitem{ref61}
	V.~Acquaviva and E.~Gawiser,
	``How to Falsify the GR+LambdaCDM Model with Galaxy Redshift Surveys,''
	Phys.\ Rev.\ D {\bf 82} (2010) 082001.
	doi:10.1103/PhysRevD.82.082001
	\bibitem{ref62}
	R.~Myrzakulov and M.~Shahalam,
	``Statefinder hierarchy of bimetric and galileon models for concordance cosmology,''
	JCAP {\bf 1310} (2013) 047.
	doi:10.1088/1475-7516/2013/10/047
	\bibitem{ref63}
	J.~Li and R.~Yang {\it et al.},
	``Discriminating dark energy models by using the statefinder hierarchy and the growth rate of matter perturbations,''
	JCAP {\bf 1412} (2014) 043.
	doi:10.1088/1475-7516/2014/12/043
	\bibitem{ref64}
	Y.~Hu and M.~Li {\it et al.},
	``Impacts of different SNLS3 light-curve fitters on cosmological consequences of interacting dark energy models,''
	Astron.\ Astrophys.\  {\bf 592} (2016) A101.
	doi:10.1051/0004-6361/201526946	
	
	\bibitem{ref65}
	A.~Mukherjee, N.~Paul and H.~K.~Jassal,
	``Constraining the dark energy statefinder hierarchy in a kinematic approach,''
	JCAP {\bf 1901} (2019) 005
	doi:10.1088/1475-7516/2019/01/005
	\bibitem{ref66}
 J.~L.~Cui, L.~Yin, L.~F.~Wang, Y.~H.~Li and X.~Zhang,
``A closer look at interacting dark energy with statefinder hierarchy and growth rate of structure,''
JCAP {\bf 1509} (2015) 024
doi:10.1088/1475-7516/2015/09/024
	
	\bibitem{ref67}
	 L.~Zhou and S.~Wang,
	``Diagnosing $\Lambda$HDE model with statefinder hierarchy and fractional growth parameter,''
	Sci.\ China Phys.\ Mech.\ Astron.\  {\bf 59} (2016) no.7,  670411
	doi:10.1007/s11433-016-0038-9
	
	\bibitem{ref68}
	
	Majumdar, A., Chattopadhyay, S., "A study of modified holographic Ricci dark energy in the framework of f (T) modified gravity and its statefinder hierarchy". Can. J. Phys. {\bf97}(5) ( 2018) 477-486.
	
	\bibitem{ref69}
	Zhao, Z., Wang, S.,  "Diagnosing holographic type dark energy models with the Statefinder hierarchy, composite null diagnostic and $\omega_ {D}-\omega_ {D}^{'} $ pair". China Phys. Mech. Astron. {\bf61}(3) (2018) 039811.
	
	\bibitem{ref70}
	Yu, F., et al.,  "Statefinder hierarchy exploration of the extended Ricci dark energy".  Eur. Phys. J. C {\bf75}(6)(2015)274.
	
	

	

	


	\bibitem{ref71}
	Mukherjee, A., Paul, N., Jassal, H. K., 2019. Constraining the dark energy statefinder hierarchy in a kinematic approach.  JCAP 2019(01), 005.
	
	

	
	\bibitem{ref72}
	Zhang, N., et al. 2019. Diagnosing the interacting Tsallis Holographic Dark Energy models. arXiv preprint arXiv:1905.04299.
	
	\bibitem{ref73}
	Srivastava, V. and Sharma.U. K.,
	"Statefinder hierarchy for Tsallis holographic dark en-
	ergy." New Astronomy (2020): 101380.
	
	
	
\bibitem {ref75}
A.R$\acute{e}$nyi , in Proceedings of the 4th Berkely Symposium on Mathematics, Statistics and Probability (University California
Press, Berkeley, CA, 1961) pp. 547–561.
\bibitem{ref75a}
V.~Sahni and A.~Shafieloo {\it et al.},
``Two new diagnostics of dark energy,''
Phys.\ Rev.\ D {\bf 78} (2008) 103502.
doi:10.1103/PhysRevD.78.103502

	\bibitem{ref76}
V.~Acquaviva and A.~Hajian {\it et al.},
``Next Generation Redshift Surveys and the Origin of Cosmic Acceleration,''
Phys.\ Rev.\ D {\bf 78} (2008) 043514.
doi:10.1103/PhysRevD.78.043514
\bibitem{ref77}
V.~Acquaviva and E.~Gawiser,
``How to Falsify the GR+LambdaCDM Model with Galaxy Redshift Surveys,''
Phys.\ Rev.\ D {\bf 82} (2010) 082001.
doi:10.1103/PhysRevD.82.082001

	\bibitem{ref78}
L.~M.~Wang and P.~J.~Steinhardt,
``Cluster abundance constraints on quintessence models,''
Astrophys.\ J.\  {\bf 508} (1998) 483.
doi:10.1086/306436
	\bibitem{ref79}
E.~V.~Linder,
``Cosmic growth history and expansion history,''
Phys.\ Rev.\ D {\bf 72} (2005) 043529.
doi:10.1103/PhysRevD.72.043529	
\end{thebibliography}
\end{document}